\documentclass[twocolumn, pre, showpacs, english, preprintnumbers, amsmath, amssymb, superscriptaddress, aps,longbibliography]{revtex4-2}
\usepackage[subpreambles=true]{standalone}

\usepackage[utf8]{inputenc}
\usepackage{graphicx}
\usepackage{grffile}
\usepackage[usenames,dvipsnames]{xcolor}
\usepackage{amsmath}
\usepackage[normalem]{ulem}
\usepackage[resetlabels, labeled]{multibib}
\usepackage{appendix}

\newcites{S}{References Supplementary Materials}
\definecolor{orange}{rgb}{1,0.5,0}
\definecolor{goodgreen}{rgb}{0.1,0.5,0}
\definecolor{goodred}{rgb}{0.7,0,0}
\usepackage{lineno}
\setpagewiselinenumbers

\usepackage{tikz}
\usepackage{tocvsec2}
\usepackage{scrextend}
\usepackage{lineno}
\modulolinenumbers[5]
\makeatletter

\usepackage[colorlinks,urlcolor=goodgreen,citecolor=blue,linkcolor=goodred]{hyperref}

\usepackage{verbatim}
\usepackage{float}

\graphicspath{ {images/} }

\let\oldepsilon\epsilon \let\epsilon\varepsilon \let\varepsilon\oldepsilon

\makeatother

\usepackage{babel}

\begin{document}

\title{Weak localization at arbitrary disorder strength in systems with generic spin-dependent fields}

\newcommand{\orcid}[1]{\href{https://orcid.org/#1}{\includegraphics[width=8pt]{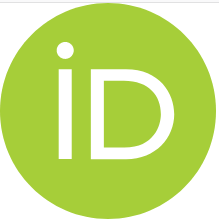}}}

\author{Alberto Hijano\orcid{0000-0002-3018-4395}}
\email{alberto.hijano@ehu.eus}
\affiliation{Centro de F\'isica de Materiales (CFM-MPC) Centro Mixto CSIC-UPV/EHU, E-20018 Donostia-San Sebasti\'an,  Spain}
\affiliation{Department of Condensed Matter Physics, University of the Basque Country UPV/EHU, 48080 Bilbao, Spain}

\author{Stefan Ili\'{c}}
\email{stefan.d.ilic@jyu.fi}
\affiliation{Centro de F\'isica de Materiales (CFM-MPC) Centro Mixto CSIC-UPV/EHU, E-20018 Donostia-San Sebasti\'an,  Spain}
\affiliation{Department of Physics and Nanoscience Center, University of Jyväskylä,
P.O. Box 35 (YFL), FI-40014 University of Jyväskylä, Finland}

\author{F. Sebasti\'{a}n Bergeret\orcid{0000-0001-6007-4878}}
\email{fs.bergeret@csic.es}
\affiliation{Centro de F\'isica de Materiales (CFM-MPC) Centro Mixto CSIC-UPV/EHU, E-20018 Donostia-San Sebasti\'an,  Spain}
\affiliation{Donostia International Physics Center (DIPC), 20018 Donostia--San Sebasti\'an, Spain}

\begin{abstract}
We present a theory of weak localization (WL) in the presence of generic spin-dependent fields, including any type of spin-orbit coupling, Zeeman fields, and non-homogeneous magnetic textures. We go beyond the usual diffusive approximation, considering systems with short-range disorder of arbitrary strength, and obtain a compact expression for the weak localization (WL) correction to the conductivity in terms of the singlet-triplet polarization operator in momentum space. The latter can be directly related to the solution of the quasiclassical Eilenberger equation for superconducting systems. This formulation presents an intuitive framework to explore how the interplay of various spin-dependent fields drives weak (anti) localization. We apply our results to study in-plane magnetoconductivity in systems with spin-orbit coupling, and in newly discovered altermagnets. Our results enable straightforward calculation of the WL conductivity at arbitrary disorder strength, which can be particularly useful for interpreting experiments on high-mobility samples.    \end{abstract}

\maketitle

\section{Introduction}

 Weak localization (WL) has been one of the central tools of mesoscopic physics \cite{bergmann1984weak, akkermans2007mesoscopic}. It enables probing quantum coherence and various symmetry breaking mechanisms in low-dimensional disordered systems by straightforward transport measurements. WL is a consequence of quantum interference — electrons traveling along closed loops interfere constructively, enhancing the probability of electrons returning to the point of origin. This leads to the reduction of the classical (Drude) conductivity. On the other hand, if strong enough spin-orbit coupling (SOC) is present in the system, the electrons interfere destructively instead, leading to the increase of Drude conductivity~\cite{hikami1980spin, iordanskii1994weak, pikus1995conduction, knap1996weak, miller2003gate, golub2005weak, glazov2006nondiffusive, glazov2009spin}. This phenomenon is known as weak anti-localization (WAL).   WAL can also be driven by mechanisms other than spin physics, for instance, sublattice pseudo-spin physics in graphene \cite{suzuura2002crossover, mccann2006weak}. 

The most common way to probe W(A)L in two-dimensional structures is to apply an out-of-plane magnetic field \cite{bergmann1984weak, akkermans2007mesoscopic}, which suppresses the interference correction to the conductivity by acting on the orbital motion of electrons. Then, by measuring the resulting magnetoconductivity and fitting it to theoretical models, one can extract microscopic parameters such as dephasing, relaxation rates, and the strength of SOC. In this work, we focus on another way to probe W(A)L, namely we study the effects on in-plane magnetic fields \cite{maekawa1981magnetoresistance, mal1997magnetoresistance, meijer2005universal, marinescu2006electron, glazov2009spin, marinescu2017cubic}.  In  2D and quasi-2D  structures, the in-plane magnetic field couples mainly to the spin of the conduction electrons via the Zeeman effect, while the orbital effect can be neglected over a large range of values of the external magnetic field. The combination of the Zeeman field with spin-dependent fields of the material, such as  SOC or magnetism, leads to distinct signatures in the in-plane magnetoconductivity. Therefore, measuring the latter provides valuable information about the material's intrinsic spin-dependent fields.

The majority of theoretical studies of W(A)L in systems with spin-dependent fields focus on the diffusive regime, e.g., in systems with SOC and Zeeman fields~\cite{iordanskii1994weak, pikus1995conduction, knap1996weak, Aleiner2001SOC, punnoose2006magnetoconductivity, marinescu2019closed, weigele2020symmetry, mal1997magnetoresistance, mal1999crystal, marinescu2006electron, marinescu2017cubic} or inhomogeneous magnetization~\cite{Loss:1993,Lyanda-Geller:1998}, meaning that the disorder scattering rate is assumed to be larger than all other energy scales of the system (except the chemical potential). This strongly disordered regime allows for simpler calculations, but in many systems, it is not justified. Namely,  advances in sample preparations allow for creating high-mobility structures, where e.g., SOC strength can be significantly larger than the disorder scattering rate. Theoretical studies beyond the usual diffusive approximation have been done in the absence of SOC in Refs.~\cite{kawabata1980theory, kawabata1984field, gasparyan1985field, cassam1994two, dyakonov1994magnetoconductance, dmitriev1997nonbackscattering, zduniak1997universal, minkov2000analysis, mcphail2004weak}, and including SOC in the works of Golub and Glazov~\cite{golub2005weak, glazov2006nondiffusive, glazov2009spin}, as well as in ballistic quantum dots~\cite{Zaitsev2005dots1,Zaitsev2005dots2}.  Interestingly, some of the signatures of the interplay of spin-dependent fields appear only beyond the diffusive approximation. For instance, in systems with Rashba SOC, magnetoconductivity is non-monotonic as a function of in-plane field, which becomes apparent only at fields comparable or larger than the disorder scattering rate \cite{glazov2009spin}. 

In this work, we go beyond existing theories and present a general theory of WAL in the presence of arbitrary spin-dependent fields at any disorder strength. Our theory includes any type of SOC, Zeeman field, as well as any type of magnetic texture. We derive a  compact expression for the quantum correction to the conductivity in terms of a singlet-triplet polarization operator in momentum space. The latter can be directly connected to the solution of the quasiclassical kinetic equation for superconducting systems, also known as the Eilenberger equation~\cite{bergeret2013singlet,ilic2020unified}. Our result provides an intuitive framework to discuss how the interplay of various spin-dependent fields influences W(A)L and enables straightforward computation of the quantum corrections to the conductivity. In particular, our findings could be useful for interpreting W(A)L measurements in high-mobility samples exhibiting any type of spin texture, both in conventional semiconducting structures, and in novel van der Waals materials and heterostructures. 

In this work, we neglect the contribution to conductivity due to electron-electron interaction (EEI) ~\cite{Altshuler-Aronov-book,Altshuler1980Interaction,Zala2001LongitudinalConductivity,Lee1985Disordered,Schwab:2003,Gornyi2004Interaction}.  EEI corrections can be similar in magnitude or larger than W(A)L corrections at low temperatures. Moreover, similarly to W(A)L, EEI corrections are sensitive to SOC ~\cite{Millis1984SOC,Minkov2012Interaction} and in-plane magnetic fields ~\cite{Zala2001Magnetoresistance}. The EEI contribution to the conductivity may be experimentally disentangled from the W(A)L by applying a sufficiently strong out-of-plane magnetic field. W(A)L correction is quickly suppressed by the field, while the EEI contribution remains mostly unaffected. Thus, the subtraction between the conductivity measured at zero and high out-of-plane field corresponds to the W(A)L conductivity alone. This procedure can be used to experimentally extract the in-plane W(A)L magnetoconductivity, which is the main quantity of interest for our work -- one should perform in-plane magnetoconductivity mesurements with and without an  out-of-plane component of the  magnetic field, and subtract the results.

The paper is organized as follows. In Sec.~\ref{Sec2}, we formulate the model for a disordered system with generic spin-dependent fields and present the main result of our work -- the compact expression for the WL conductivity at arbitrary disorder strength and in the presence of any spin-dependent fields. In this Section, we also discuss the relationship of our results to the Eilenberger equation. In Sec.~\ref{Sec3}, we present several applications of our results -- we discuss systems with various kinds of SOC subjected to Zeeman fields, and we examine the WL in newly discovered altermagnets. We conclude and summarize our results in Sec.~\ref{Sec4}.

\section{The Model and Main Results \label{Sec2}}

In Sec.~\ref{SecEff} we first present the effective Hamiltonian to describe electronic systems with generic spin-dependent fields and disorder strength. 
We then introduce  the main results of this work.  In Sec.~\ref{Sec3A}, we show the general expression for the WL correction to the conductivity. For brevity and clarity, technical details of the derivation of these results from the diagrammatic theory are omitted in the main text, and shown instead in Appendix \ref{AppMainRes}. In Sec.~\ref{Sec3C}, we discuss the relationship of our result to the quasiclassical kinetic equation. Finally, in Sec.~\ref{Sec3B}, we discuss our results in the diffusive regime, where the expression for the WL conductivity significantly simplifies.

\subsection{Effective Hamiltonian \label{SecEff}}
The effective Hamiltonian describing a  system with generic spin-dependent fields and disorder strength has the form
\begin{equation}\label{Hamiltonian}
\mathcal{H}=\xi_{\boldsymbol{p}} + b^i_{\boldsymbol{p}}\sigma_i+h^i_{\boldsymbol{p}}\sigma_i+V\; .
\end{equation}
Here, $\xi_{\boldsymbol{p}}=\boldsymbol{p}^2/(2m)-\mu$ is the kinetic energy term, where $\mu$ is the chemical potential. Spin-dependent fields can be classified as SOC terms, $b^i_{\boldsymbol{p}}$, that break the inversion symmetry, and the magnetic (Zeeman-like) terms, $h^i_{\boldsymbol{p}}$, that break the time-reversal symmetry (TRS). The SOC terms are odd under sign change of momentum  $b^i_{-{\boldsymbol{p}}}=-b^i_{\boldsymbol{p}}$, while the magnetic terms are even $h^i_{-{\boldsymbol{p}}}=h^i_{\boldsymbol{p}}$. 
We introduced the Pauli matrices in spin space $\sigma_i$, with $i=x,y,z$. The effect of disorder is captured by the random potential $V$, with an associated scattering time $\tau$. We take that disorder scattering is isotropic and spin-independent.

Throughout this work, we will employ the quasiclassical approximation, which assumes that the chemical potential is the largest energy scale in the system, excluding the Fermi energy: $\mu \gg b_{\boldsymbol{p}}^i, h_{\boldsymbol{p}}^i, \tau^{-1}$. This assumption is valid for the majority of experimentally available structures, and it allows us to analytically evaluate the energy integrals in the diagrammatic theory of WL (see Appendix \ref{AppMainRes}). Because electronic transport occurs at the Fermi level, we may approximate $b_{\boldsymbol{p}}\approx b_{p_F \boldsymbol{n}} \equiv b_{\boldsymbol{n}}$ and $h_{\boldsymbol{p}}\approx h_{p_F \boldsymbol{n}} \equiv h_{\boldsymbol{n}}$, where $p_F$ is the Fermi momentum, and $\boldsymbol{n}=\boldsymbol{p}/|\boldsymbol{p}|$ is the unit vector associated with the direction of momentum along the Fermi surface.


\subsection{WL correction to the conductivity {\label{Sec3A}}}
The central object in our theory is the angle-resolved polarization operator  $\hat{\Pi}_{\boldsymbol{n}\boldsymbol{Q}}$, written in momentum $\boldsymbol{Q}$ space
\begin{equation}
\hat{\Pi}_{\boldsymbol{n}\boldsymbol{Q}}= (1+l/l_\phi+i l \boldsymbol{n}\cdot \boldsymbol{Q}+\hat{A}_{\boldsymbol{n}})^{-1}.
\label{eq:Pol}
\end{equation}
Here, $l=v_F \tau$ is the mean free path, and $l_\phi=v_F \tau_\phi\gg l$ is the decoherence length, with $\tau_\phi$ being the electron decoherence time. The matrix $\hat{A}_{\boldsymbol{n}}$ accounts for the  spin-dependent fields: 
\begin{equation}
\hat{A}_{\boldsymbol{n}}= 2\tau
\begin{pmatrix}
0 & b^z_{\boldsymbol{n}} & -b^y_{\boldsymbol{n}} & -i h^x_{\boldsymbol{n}}\\ 
-b^z_{\boldsymbol{n}} & 0 & b^x_{\boldsymbol{n}} &-i  h^y_{\boldsymbol{n}} \\
b^y_{\boldsymbol{n}} & -b^x_{\boldsymbol{n}} & 0 & -i h^z_{\boldsymbol{n}}\\
-i h^x_{\boldsymbol{n}} & -i h^y_{\boldsymbol{n}} & -i h^z_{\boldsymbol{n}} & 0
\end{pmatrix}.
\label{eq:An}
\end{equation}
Here, the symbol $\, \hat{} \,$ denotes that the object is a $4\times 4$ matrix in the singlet-triplet space, spanning the three triplets and the singlet built from spin states of interfering pairs of electrons. We will label the components of a matrix $\hat{K}$ as $\hat{K}_{ij}$, with $i,j=x,y,z,0$, where indices $x,y,z$ denote the three triplets, while the index $0$ denotes the singlet. From the form of Eq.~\eqref{eq:An}, we can see what is the role of different spin-dependent fields:  SOC terms $b_{\boldsymbol{n}}^i$  couple different triplets (causing triplet precession and relaxation),  while the magnetic terms $h_{\boldsymbol{n}}^i$ couple the singlet to the triplets.

We proceed to define another important object in the theory of WL -- the Cooperon $\hat{C}_{\boldsymbol{Q}}$, which gives the amplitude of quantum interference of two electrons traveling in opposite directions. It can be written as
\begin{equation}
\hat{C}^{-1}_{\boldsymbol{Q}}=2\pi \nu \tau[1-\hat{\Pi}_{\boldsymbol{Q}}].
\label{Eq:Coop}
\end{equation}
Here, $\hat{\Pi}_{\boldsymbol{Q}}=\langle \hat{\Pi}_{\boldsymbol{n}\boldsymbol{Q}}\rangle$, where $\langle ... \rangle$ represents averaging over the direction $\boldsymbol{n}$, and $\nu$ is the density of states at the Fermi level.

 We can finally write the quantum  correction to the conductivity in a very compact form:
\begin{equation}
\delta\sigma=\frac{e^2}{2\pi}\int \frac{d^d\boldsymbol{Q}}{(2\pi)^d}\text{Tr}[\hat{C}^{(a)}_{\boldsymbol{Q}}\hat{W}^{(a)}_{\boldsymbol{Q}}
+\hat{C}^{(b)}_{\boldsymbol{Q}}\hat{W}^{(b)}_{\boldsymbol{Q}}
],
\label{eq:sigma}
\end{equation}
where $d$ is the dimensionality of the system. The first and second terms in Eq.~\eqref{eq:sigma} correspond to the backscattering and non-backscattering contributions to the WL conductivity, respectively~\cite{dmitriev1997nonbackscattering}. Formally, they come from the bare and dressed Hikami boxes in the diagrammatic theory, respectively (see Appendix~\ref{AppMainRes}). The matrices $\hat{W}_{\boldsymbol{Q}}^{(a)}$ and $\hat{W}_{\boldsymbol{Q}}^{(b)}$ are the Cooperon weight factors, which are given as
\begin{align}
\label{eq:Wa}
&\hat{W}_{\boldsymbol{Q}}^{(a)}= \hat{W}_0 \left\langle n_x^2 \hat{\Pi}_{\boldsymbol{n\boldsymbol{Q}}}^\dagger\hat{\Pi}_{\boldsymbol{n\boldsymbol{Q}}}\right\rangle  \\
\label{eq:Wb}
&\hat{W}_{\boldsymbol{Q}}^{(b)}=-\hat{W}_0 \left \langle n_x \hat{\Pi}_{\boldsymbol{n\boldsymbol{Q}}}^\dagger \right\rangle
\left \langle n_x \hat{\Pi}_{\boldsymbol{n\boldsymbol{Q}}} \right\rangle. 
\end{align}
Here, the current is taken to flow along the $x$-direction, and $n_x$ is the component of the vector $\boldsymbol{n}$ in this direction. We introduced the matrix $\hat{W}_0=4 \pi \nu \tau^3 v_F^2 \, \text{diag}(-1,-1,-1,1)$, which ensures that the singlet and triplet contributions enter with opposite signs in the WL conductivity \cite{bergmann1984weak, akkermans2007mesoscopic}. The objects $\hat{C}^{(a)}_{\boldsymbol{Q}}$  and $\hat{C}^{(b)}_{\boldsymbol{Q}}$ are the reduced Cooperons, $\hat{C}^{(a)}_{\boldsymbol{Q}}=\hat{C}_{\boldsymbol{Q}} \hat{\Pi}_{\boldsymbol{Q}}^2$ and $\hat{C}^{(b)}_{\boldsymbol{Q}}=\hat{C}_{\boldsymbol{Q}} \hat{\Pi}_{\boldsymbol{Q}}$. 
Here, we consider only the contributions of closed paths with three or more impurity collisions to the quantum interference. This is the usual approach when dealing with systems beyond diffusive approximation ~\cite{gasparyan1985field, cassam1994two, dyakonov1994magnetoconductance, dmitriev1997nonbackscattering, zduniak1997universal, golub2005weak,glazov2006nondiffusive,glazov2009spin}, where the divergent contributions of paths with single or two colissions need to be removed.

Equation~\eqref{eq:sigma} is one of the main results of our work. We see that the quantum correction to conductivity can be expressed in terms of the polarization operator $\hat{\Pi}_{\boldsymbol{n Q}}$, Eq.~\eqref{eq:Pol}, which enters the Cooperon and the weight matrices $\hat{W}^{(a,b)}_{\boldsymbol{Q}}$ through Eqs. (\ref{Eq:Coop}) and (\ref{eq:Wa}-\ref{eq:Wb}), respectively. As we discuss in Sec.~\ref{Sec3C}, the polarization operator can be directly obtained also from the Eilenberger equation.

The conductivity is obtained by evaluating the momentum integral in Eq.~\eqref{eq:sigma}. Here, having a finite $\tau_\phi$ is required to remove the infrared divergence of the integral. Note that it is not necessary to impose an upper cutoff for the $Q$-integration. This is in contrast to the standard approach within the diffusive approximation, where it is customary to impose the cutoff $Q_{max}\sim 1/(v_F \tau)$ to ensure convergence \cite{bergmann1984weak, akkermans2007mesoscopic}.  

The results of this section hold for any system dimensionality $d$, but in the following we will focus on 2D systems, so that $\boldsymbol{p}=(p_x,p_y)$ and $\boldsymbol{n}=(n_x,n_y)$.

%
%
\begin{figure*}[!t]
  \includegraphics[width=0.99\textwidth]{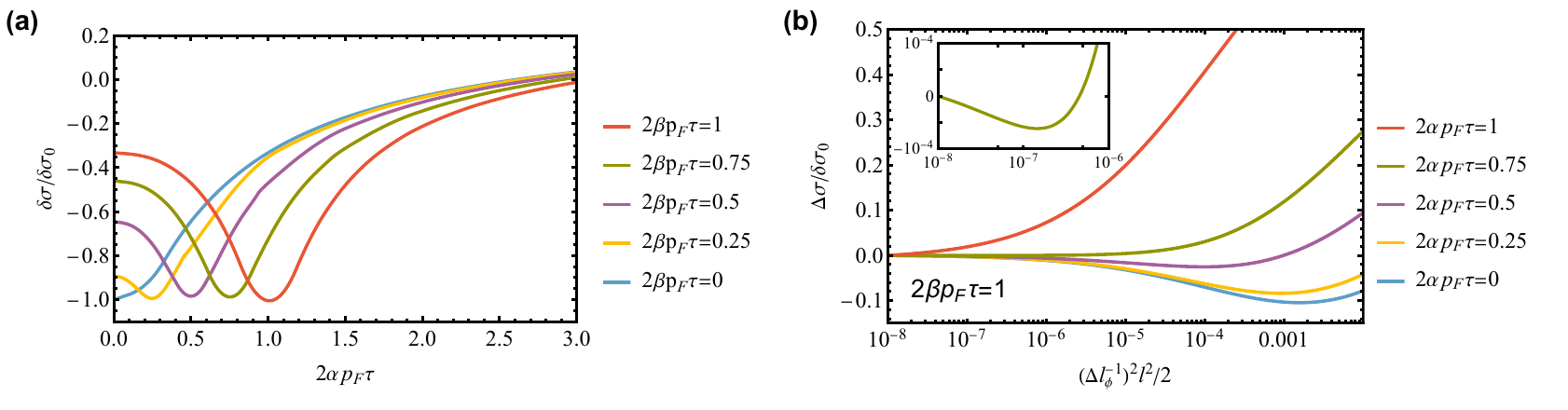}
    \caption{(a) Interference correction to the conductivity $\delta\sigma$ as a function of the Rashba $\alpha$ and Dresselhaus $\beta$ SOC strengths. The ratio between the decoherence length and the mean free path is fixed to $l_\phi/l=100$. (b) WL conductivity $\Delta\sigma=\delta\sigma(l_\phi^{-1}+\Delta l_\phi^{-1})-\delta\sigma(l_\phi^{-1})$ as a function of the decoherence length $\Delta l_\phi^{-1}$ and $\alpha$, for $2\beta p_F \tau=1$. The inset in panel (b) is a zoom-in of the green curve around its minimum.}
    \label{Fig:conductance_vs_Rashba_magnetic}
\end{figure*}
%
%

\subsection{Relation to the Eilenberger equation \label{Sec3C}}
In this section, we connect the WL conductivity to the quasiclassical kinetic equation for superconducting systems, also known as Eilenberger equation~\cite{bergeret2013singlet,ilic2020unified}. 
This equation is a widely used tool for describing superconductors at arbitrary degree of disorder. It has also been formulated in the presence of spin-dependent fields \cite{bergeret2013singlet}, where triplet superconducting correlations come into play. As we will discuss in the following, the polarization operator and Cooperons entering the WL formalism can be directly obtained from the linear Eilenberger equation. Moreover, if the TRS is not broken, the linear Eilenberger equation is formally identical to the kinetic equation for coupled spin-charge transport in the normal state. The linear Eilenberger equation therefore presents a unified framework to describe different phenomena driven by spin-dependent fields: triplet superconductivity, spin transport, and W(A)L, as discussed in Ref.~\cite{ilic2020unified}.

For a system described by the Hamiltonian ~\eqref{Hamiltonian}, the linear Eilenberger equation in momentum space reads \cite{bergeret2013singlet}
\begin{equation}\label{Eilenberger}
(1+l/l_\phi+i\boldsymbol{n}\cdot \boldsymbol{Q}l)f+i \tau [b_{\boldsymbol{n}}^i\sigma_i, f]-i\tau \{h^i_{\boldsymbol{n}}, f \} = \langle f \rangle.
\end{equation}
Here, we omit the superconducting source term which is not relevant for our discussion. The anomalous Greeen's function $f=f_i\sigma_i$ ($i=0,x,y,z$)  captures the singlet $(i=0)$ and triplet $(i=x,y,z)$ superconducting correlations. $\sigma_0$ is the unit matrix. The above equation can be rewritten as 
\begin{equation}
\vec{f}=\hat{\Pi}_{\boldsymbol{n Q}}\langle \vec{f} \rangle\; ,
\end{equation}
where $\vec{f}=(f_x,f_y,f_z,f_0)^T$. Therefore, the polarization operator appears directly in the linear Eilenberger equation. Moreover, after averaging over momentum direction, we have
\begin{equation}
( 1-\langle \hat{\Pi}_{\boldsymbol{nQ}} \rangle )\langle \vec{f}\rangle =0\; ,
\end{equation}
which becomes 
\begin{equation}
\hat{C}_{\boldsymbol{Q}}^{-1} \langle \vec{f}\rangle=0. 
\end{equation}
Therefore, the Cooperon is the resolvent of the linear Eilenberger equation. We see that there is a close connection between superconductivity and WL. It comes from the fact that the superconducting (particle-hole) correlations of the linear Eilenberger equation are equivalent to maximally crossed diagrams from the theory of WL (see Appendix \ref{AppMainRes}). The Eilenberger equation approach provides a simple and physically transparent picture to compute the Cooperon for arbitrary strength of disorder, which allows for analytical solutions in many cases.

\subsection{Diffusive limit \label{Sec3B}}
As mentioned in the introduction, most of the existing theoretical studies on WL \cite{iordanskii1994weak, pikus1995conduction, knap1996weak, punnoose2006magnetoconductivity, marinescu2019closed, weigele2020symmetry, mal1997magnetoresistance, marinescu2006electron, marinescu2017cubic} focus on the diffusive limit where $lQ, \tau b^i_{\boldsymbol{n}}, \tau h^i_{\boldsymbol{n}} \ll 1$. In this regime,  the expressions \eqref{Eq:Coop}-\eqref{eq:Wb} further simplify. Namely, we can find the polarization operator as
\begin{equation}
1-\hat{\Pi}_{\boldsymbol{Q}}\approx l/l_\phi+D Q^2 \tau+\langle \hat{A}_{\boldsymbol{n}}\rangle- 2i  \langle l \boldsymbol{Q}\cdot \boldsymbol{n} \hat{A}_{\boldsymbol{n}}   \rangle - \langle \hat{A}_{\boldsymbol{n}}^2\rangle,
\label{Eq:PolDif}
\end{equation}
and the Cooperon is readily obtained using Eq.~\eqref{Eq:Coop}. Here, we introduced the diffusion constant $D=\frac{1}{d}v_F^2\tau$. Only the magnetic fields $h_{\boldsymbol{n}}^i$ contribute to the term $\langle \hat{A}_{\boldsymbol{n}}\rangle$, which captures singlet-triplet coupling. Similarly, only the SOC fields $b_{\boldsymbol{n}}^i$ contribute to the term $2i  \langle l \boldsymbol{Q}\cdot \boldsymbol{n} \hat{A}_{\boldsymbol{n}}   \rangle$,  capturing triplet precession, that is, conversion of one triplet to another. Finally, the term  $\langle \hat{A}_{\boldsymbol{n}}^2\rangle$ describes the singlet and triplet relaxation, due to SOC and magnetic fields. The Cooperon weight factors become
\begin{equation}
\hat{W}^{(a)}=\frac{1}{d}\hat{W}_0,
\end{equation}
while $\hat{W}^{(b)}=0$.
The WL correction to the conductivity can now be written as
\begin{equation}
\delta\sigma=\frac{e^2}{2\pi d}\frac{1}{2\pi \nu \tau}\int \frac{d^d\boldsymbol{Q}}{(2\pi)^d}\text{Tr}[(1-\hat{\Pi}_{\boldsymbol{Q}})^{-1}\hat{W}_0
].
\label{eq:sigmaDif}
\end{equation}
Importantly, as mentioned in Sec.~\ref{Sec3A}, it is necessary to impose an upper cutoff for momentum integration $Q_{max}\sim 1/(v_F \tau)$ to ensure convergence of Eq.~\eqref{eq:sigmaDif}. 

The expressions \eqref{Eq:PolDif}-\eqref{eq:sigmaDif} generalize previous theoretical works valid in the diffusive limit \cite{iordanskii1994weak, pikus1995conduction, knap1996weak, punnoose2006magnetoconductivity, marinescu2019closed, weigele2020symmetry, mal1997magnetoresistance, marinescu2006electron, marinescu2017cubic} by accounting for arbitrary spin-dependent fields. As an example,  in  Appendix~\ref{AppDif}, we reproduce the polarization operator in the diffusive limit for the well studied case of a 2D system with Rashba and Dresselhaus SOC using Eq.~\eqref{Eq:PolDif}.

%
%
\begin{figure*}[!t]
  \includegraphics[width=0.99\textwidth]{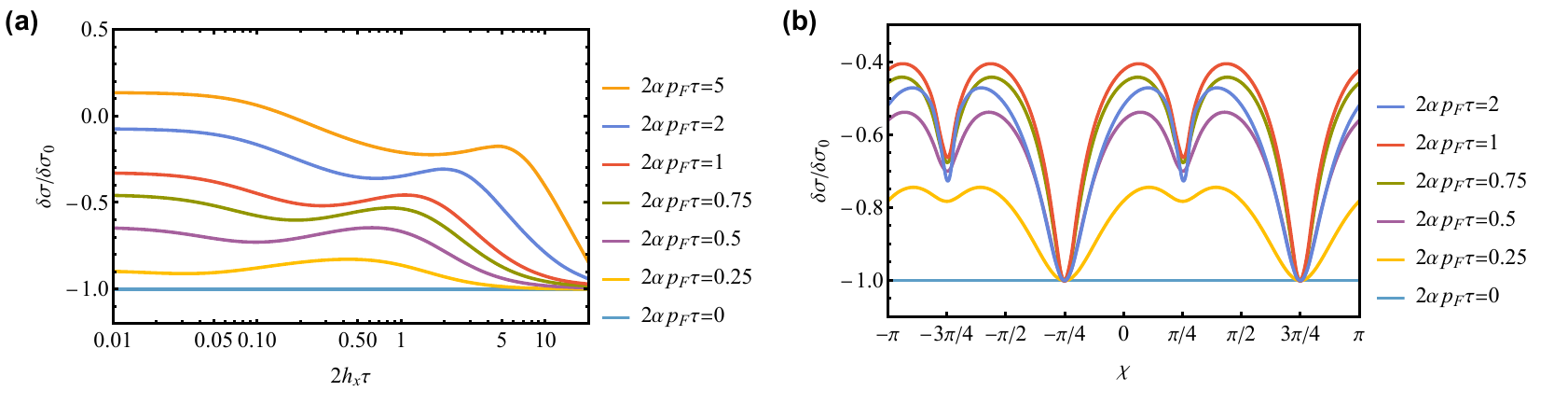}
    \caption{Interference correction to the conductivity as a function of the (a) magnitude of the external in-plane field, $h_x$ and (b) its direction, characterized by the angle $\chi$ with respect to the current. In panel (a) $\beta=0$ and $h_y=0$, and in panel (b) $\beta=\alpha$ and $2h\tau=1$.}
    \label{Fig:conductance_vs_Zeeman_gamma}
\end{figure*}
%
%

\section{Applications \label{Sec3}}
In this Section we apply the formalism introduced in Sec.~\ref{Sec2} to compute the conductivity correction to different systems with spin-dependent fields. In Sec.~\ref{sec:RashbaDresselhaus} we first address the well studied 2D systems with Rashba and Dresselhaus SOC. In addition to reproducing known results, we propose a new way to probe these types of SOC in high-mobility systems by measuring in-plane WL magnetoconductivity for different orientations of the field. In the rest of the Section, we address systems with more exotic spin-field types and propose WL measurements to probe them: materials with the so-called Weyl-type SOC in Sec.~\ref{sec:Weyl}, and newly discovered altermagnets in Sec.~\ref{sec:Altermagnet}.

\subsection{Systems with Rashba and Dresselhaus SOC\label{sec:RashbaDresselhaus}}

In this section, we focus on systems with Rashba and Dresselhaus SOC (both linear and cubic). The Rashba SOC arises due to the lack of structural inversion symmetry~\cite{Rashba}, while the Dresselhaus SOC arises in materials with broken lattice inversion symmetry~\cite{Dresselhaus}. Crystals with SOC show a variety of unique transport properties that have extensively been exploited in the field of spintronics~\cite{Zutic2004spintronics}. Their applications include spin field-effect transistors~\cite{datta-das-90,Koo_2009,datta-das-15}, spin filtering~\cite{Aharony_2008}, and spin-based platforms for topological quantum computing~\cite{Fu-Kane-09}, among others. SOC has extensively been studied in the context of WAL. The presence of SOC hinders the backscattering due to the constructive electron interference, adding a positive correction to the conductivity~\cite{hikami1980spin,Bergmann1982WAL}. 


The spin-dependent fields $b_{\boldsymbol{n}}^i$ for a systems with Rashba SOC ($\alpha)$, and linear ($\beta)$ and cubic ($\beta_3)$ Dresselhaus SOC are
\begin{align}
&b_{\boldsymbol{n}}^x=\alpha p_F n_y+\beta p_F n_x+\beta_3 p_F^3 (n_x^3-3n_xn_y^2), \nonumber \\
&b_{\boldsymbol{n}}^y=-\alpha p_F n_x-\beta p_F n_y-\beta_3 p_F^3 (n_y^3-3n_y n_x^2).
\label{Eq:Rashba}
\end{align}
The momentum-independent in-plane Zeeman fields are captured by the terms
\begin{equation}
h_{\boldsymbol{n}}^x=h_x, \qquad  h_{\boldsymbol{n}}^y=h_y. 
\label{eq:Zeeman}
\end{equation}

In Fig.~\ref{Fig:conductance_vs_Rashba_magnetic}(a) we show the WL correction to the Drude conductivity for different values of the Rashba and Dresselhaus SOCs. The conductivity is normalized with $\delta \sigma_0=\frac{e^2}{2\pi^2}\ln{\frac{\tau_\phi}{2\tau}}$, so that in the absence of spin-dependent fields the  WL correction is $\delta \sigma=-\delta \sigma_0$ ~\cite{glazov2006nondiffusive,glazov2009spin}. The decoherence length is set to $l_\phi=100l$ in all plots, except for Fig. 1(b). In the absence of Dresselhaus SOC (blue curve), the conductivity increases with increasing strength of the Rashba SOC, in agreement with the known results. The SOC induces a momentum-dependent spin precession around the \textit{effective SOC field} $\boldsymbol{b}_{\boldsymbol{n}}=(b_{\boldsymbol{n}}^x,b_{\boldsymbol{n}}^y,b_{\boldsymbol{n}}^z)$, modifying the interference of electron waves and leading to WAL~\cite{Bergmann1982WAL,knap1996weak,Wenk2010antilocalization}. The conductivity correction saturates at $\delta\sigma\approx \frac{1}{2}\delta \sigma_0-0.2 \frac{e^2}{2\pi^2}$ \cite{glazov2006nondiffusive} for infinite SOC ($\alpha \to \infty$). The WL conductivity correction consists of one (positive) singlet and three (negative) triplet contributions. The three triplet channels are suppressed by the SOC, so the conductivity correction is given by the remaining singlet channel \cite{hikami1980spin}.

For finite Rashba and Dressehaus SOC, the conductivity decreases with increasing $\alpha$ until it reaches a minimum at $\alpha=\beta$. The interplay between the two types of SOC is substractive~\cite{pikus1995conduction} -- when they are of equal strengths, the effective SOC field decouples from momentum and points always in the same direction. This effect is known as persistent spin helix, as the spin precession only depends on the distance travelled along the $(1/\sqrt{2},1/\sqrt{2},0)$ direction~\cite{Bernevig2006spinhelix}. Then,  the SU(2) phase associated to the spin precession will be zero for all closed paths contributing to the WL. Thus, the conductivity for $\alpha=\beta$ is equal to the spin-field-free one $\delta\sigma=-\delta\sigma_0$.

In experiments measuring in-plane WL magnetoconductivity, the orbital effects of the magnetic field may still play an important role. Firstly, there might be a finite out-of-plane component of the field due to misalignment. Secondly, the in-plane component  of the field may also act on the orbital motion of electrons in 2D films due to the finite thickness of the film. Both effects can have a significant impact on the WL conductivity, especially at high fields comparable to SOC strength. However, exact treatment of orbital effects is a computationally demanding task (especially beyond the diffusive approximation), for which the real-space formulation of the WL theory ~\cite{golub2005weak, glazov2006nondiffusive, glazov2009spin}  is better suited than our momentum space approach. Instead of exact computations, the orbital effects may be qualitatively captured by renormalizing the decoherence length, $l_\phi^{-1}\to l_\phi^{-1}+\Delta l_\phi^{-1}$. Specifically, $\Delta l_\phi^{-1} \propto (2eB_\perp)^{1/2}$ \cite{Lyanda-Geller1998Quantum, marinescu2019closed}  captures a weak out-of-plane component of the field $B_\perp$, while  $\Delta l_\phi^{-1} \propto l B_{||}^2 d^2$ captures the orbital effect of an in-plane field $B_{||}$ for a 2D film of thickness $d$ \cite{mal1999crystal}.  In Fig.~\ref{Fig:conductance_vs_Rashba_magnetic}(b) we plot the correction to the conductivity as a function of  $\Delta l_\phi^{-1}$ for a system with Rashba and Dresselhaus SOC.  Importantly, the effects analyzed in this manuscript are only measurable if the orbital effects are sufficiently weak, so that the signatures of the interplay between spin-dependent fields in the WL conductance are not washed out.





In Fig.~\ref{Fig:conductance_vs_Zeeman_gamma}(a) we show the weak-localization correction to the conductivity due to a Zeeman field for different Rashba SOC strengths. In the absence of SOC, the Zeeman field is the only spin-dependent field. The system has a definite spin quantization axis, so working on the appropriate spin basis the SU(2) phase acquired by all paths contributing to the WL conductivity will be trivial. The SOC increases the conductivity value due to its antilocalization effect, even though very strong Zeeman fields completely overcome the SOC, restoring the usual $\delta\sigma=-\delta \sigma_0$ correction. Note that the conductivity curves have a bump close to $h_x\approx \alpha p_F$. This non-monotonous behavior appears only beyond the diffusive approximation, as first predicted in Ref.~\cite{glazov2009spin}.

In Fig.~\ref{Fig:conductance_vs_Zeeman_gamma}(b) we study the conductivity dependence on the Zeeman field orientation $\chi$, where $\chi$ is measured from the current direction. Here we choose $\alpha=\beta$, where the dependence on angle $\chi$ is maximized due to the formation of the persistent spin helix. For $\chi=-\pi/4$ and $\chi=3\pi/4$, the Zeeman field is collinear with the SOC spin-field, so $\delta\sigma=-\delta \sigma_0$. The conductivity curves are symmetric with respect to both values of $\chi$, since both Zeeman field configurations are related through the spin rotation $\widehat{R}_{\hat{\boldsymbol{n}}_-}(\pi)$, where $\hat{\boldsymbol{n}}_-=(1/\sqrt{2},-1/\sqrt{2},0)$, which leaves the SOC field $\boldsymbol{b}_{\boldsymbol{p}}$ unchanged (see Appendix \ref{AppRot}). The conductivity shows two shallower minima at $\chi=\pi/4$ and $\chi=-3\pi/4$, with the depth of the minima increasing with increasing SOC strength. This feature is absent in the diffusive regime~\cite{mal1997magnetoresistance,mal1999crystal}: in the dirty metal limit the spin-dependent fields are weak compared to the scattering frequency $\alpha p_F \tau,\alpha p_F \tau,h \tau \ll 1$, so the SOC is not strong enough to develop a minimum.

\begin{figure}[!t]
  \includegraphics[width=0.99\columnwidth]{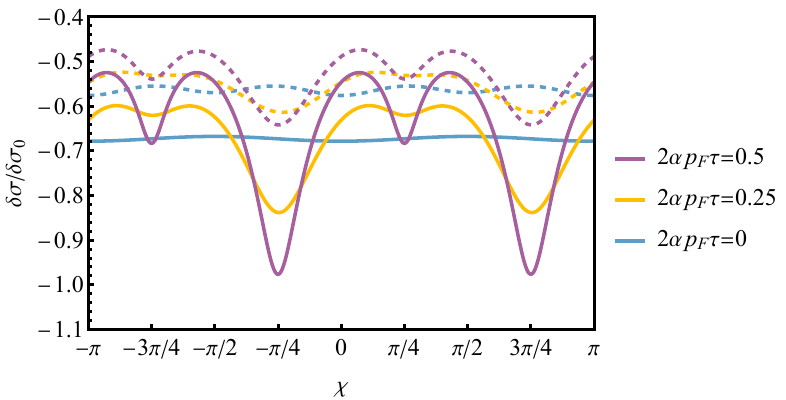}
\caption{Interference correction to the conductivity as a function of $\chi$ and $\alpha$. The linear Dresselhaus SOC term is set to $2\beta p_F \tau=0.5$, the cubic term to $2\beta_3 p_F^3 \tau=0$ (solid) or $2\beta_3 p_F^3 \tau=0.25$ (dashed), and $2h\tau=1$.}\label{Fig:conductance_vs_gamma_cubic}
\end{figure}

In Fig.~\ref{Fig:conductance_vs_gamma_cubic}, we investigate different ratios of $\alpha$ and $\beta$ and show the influence of the cubic Dresselhaus term $\beta_3$. The solid lines correspond to $\beta _3=0$ and the dashed lines to $2\beta_3 p_F^3=0.25$. We see that the shape of $\delta \sigma (\chi)$ curves strongly depends on the ratio $\alpha/\beta$, while the cubic Dresselhaus term generally contributes to the WAL by increasing the value of the conductivity. In addition, the cubic Dresselhaus term breaks the symmetry around the $\chi=\pi/4,-3\pi/4$ angles present in the persistent spin helix regime ($\alpha=\beta$, purple lines). The cubic Dressenhaus SOC creates a small difference in the height of maxima, with a stronger cubic term leading to a greater asymmetry.

%
%
%
%

\subsection{ Systems with Weyl-type SOC \label{sec:Weyl}}
While Rashba and Dresselhaus SOCs have been extensively studied over the last decades, there are other types of linear-in-momentum SOCs that can be distinguished. In quasi-2D materials, the momentum $\boldsymbol{p}$ has two components, while the spin can point along any spatial direction. Focusing on SOCs with inplane spin-momentum locking, we can differentiate four linearly independent types of SOC: Rashba type SOC $p_y\sigma_x-p_x\sigma_y$, two Dresselhaus type SOCs $p_x\sigma_x-p_y\sigma_y$ and $p_y\sigma_x+p_x\sigma_y$; and Weyl-type SOC $p_x\sigma_x+p_y\sigma_y$. While Rashba and Dresselhaus SOCs may arise in achiral non-centrosymmetric crystals, Weyl-type SOC requires chirality. Chiral topological semimetals that host single- and multifold band crossings have recently been predicted to show this kind of SOC. Some material candidates to host Weyl-type SOC are YSb$_2$~\cite{Chang2018Weyl} and PtGa~\cite{Lin2022Weyl,Krieger2022Weyl}.

\begin{figure}[!t]
  \includegraphics[width=0.99\columnwidth]{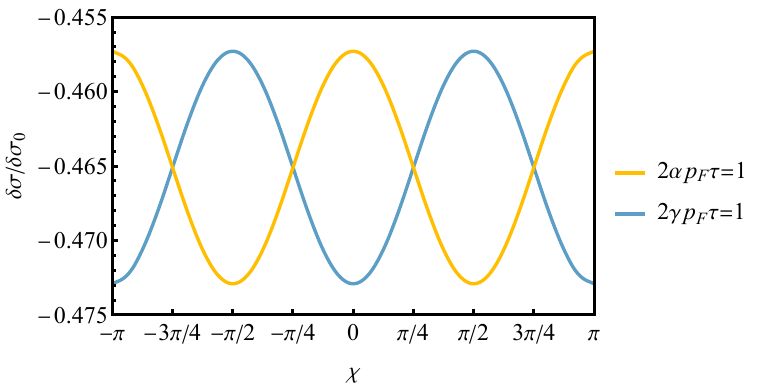}
\caption{Interference correction to the conductivity as a function of $\chi$ for a system with Rashba (yellow) or Weyl-type (blue) SOC. The strength of the Zeeman field is $2h\tau=1$.}\label{Fig:conductance_vs_gamma_parallel}
\end{figure}

In this subsection, we study the WL conductivity of materials with Weyl-type parallel SOC in the presence of in-plane Zeeman fields, analyze their relation to Rashba and Dresselhaus SOCs, and propose conductivity measurements that allow to distinguish all SOC types. The spin-dependent fields for materials with Weyl-type SOC read
\begin{equation}
b_{\boldsymbol{n}}^x =\gamma p_F n_x, \qquad b_{\boldsymbol{n}}^y=\gamma p_F n_y,
\label{eq:Weyl}
\end{equation} 
and we consider isotropic Zeeman terms introduced in Eq.~\eqref{eq:Zeeman}. As seen in Eq.~\eqref{eq:Weyl}, the effective SOC field $\boldsymbol{b}_{\boldsymbol{n}}$ is parallel to the momentum direction. This is in contrast to the Rashba effective SOC field, which is perpendicular to the motion direction. As proven in Appendix~\ref{AppRot}, systems where the spin-field interactions are related through spin symmetry operations have equivalent WL correction. Rotating $\boldsymbol{b}_{\boldsymbol{n}}$ and $\boldsymbol{h}_{\boldsymbol{n}}=(h_x,h_y,0)$ by $-\pi/2$ around the $z$ axis, we obtain that the WL conductivity for a system with Weyl-type SOC is equivalent to a system with Rashba SOC of strength $\gamma$ and Zeeman field $\boldsymbol{h}=(h_y,-h_x,0)$ (see third entry in Table~\ref{table:spin_symmetries}). One way to identify Weyl-type SOC interaction and distinguish it from Rashba type SOC is to study the dependence of the conductivity on the direction of the in-plane Zeeman field with respect to the current $\chi$. As shown in Fig.~\ref{Fig:conductance_vs_gamma_parallel}, the in-plane magnetoconductivity for a system with a single type of SOC is anisotropic. Note that the magnetoconductivity oscillations are lost within the conventional diffusive approximation, where a combination of different types of SOC are needed to obtain an in-plane anisotropy~\cite{mal1997magnetoresistance,mal1999crystal}. The conductivity for a system with Rashba SOC shows a maximum for Zeeman field orientations $\chi=0$ and $\chi=\pi$, while the conductivity for a system with Weyl-type SOC shows minima for these orientations. 


%
%
%
%

Both Rashba and Weyl-type SOCs are isotropic with respect to inplane rotations. This is in contrast to the Dresselhaus SOC. One simple way to discriminate Weyl-type and Dresselhaus SOC is to measure the longitudinal conductivity along different directions. In the former, the conductivity vs $\chi$ curves will be identical for any current direction. For the Dresselhaus SOC, the conductivity curves will differ for different current directions. For instance, using the fourth symmetry operation in Table~\ref{table:spin_symmetries}, we obtain that the conductivity along the $x$-direction for a material with Dresselhaus SOC is equivalent to a material with Weyl-type SOC and $y$-mirrored Zeeman field. For $2\beta p_F \tau=1$ and $2 h \tau=1$, the conductivity pattern is equivalent to the blue curve in Fig.~\ref{Fig:conductance_vs_gamma_parallel}. However, if the conductivity is measured along the $(1/\sqrt{2},1/\sqrt{2},0)$ direction, using the second symmetry operation in Table~\ref{table:spin_symmetries} we obtain that the conductivity is equivalent to the yellow curve in Fig.~\ref{Fig:conductance_vs_gamma_parallel} (Rashba case). Therefore, it is possible to unequivocally discriminate Weyl-type SOC from the more common Rashba and Dresselhaus SOCs by performing conductivity measurements along two directions.

\subsection{ Altermagnets \label{sec:Altermagnet}}
Traditionally, materials with magnetic ordering have been classified into ferromagnets and antiferromagnets. In the former, the magnetic moments align leading to a macroscopic magnetization. The TRS of the electronic bands is broken on ferromagnets, which results in the spin-splitting of the energy bands. In antiferromagnets, the atomic magnetic moments order in such a way that the material does not show a net magnetization. Despite the magnetic ordering, the energy bands of antiferromagnets are spin-degenerate, as in nonmagnetic materials. Recently, a new kind of materials combining ferromagnetic and antiferromagnetic properties, known as altermagnets, have been theoretically predicted~\cite{Yuan2020altermagnetism}.

Altermagnets consist of opposite-spin sublattices with zero net magnetization but spin-split band structures. Unlike materials with SOC, the spin splitting of the energy bands in altermagnets does not arise due to relativistic effects, so they do not require inversion symmetry or high atomic number materials. Moreover, due to its non-relativistic origin, the spin-splitting in altermagnets~\cite{Ahn2019RuO2,Smejkal2020RuO2,Yuan2020altermagnetism} is much stronger than in materials with SOC~\cite{Ishizaka2011giantspinsplitting}. Materials that show this kind of magnetism include crystals with rutile (P$4_2$/mnm) structure such as RuO$_2$~\cite{Ahn2019RuO2,Smejkal2020RuO2}, FeF$_2$~\cite{Lopez-Moreno2012FeF2} and MnF$_2$~\cite{Lopez-Moreno2016MnF2,Yuan2020altermagnetism}, as well as several crystals with NiAs-type structure (P$6_3$/mmc), such as MnTe~\cite{Mazin2023MnTe} and EuIn$_2$As$_2$~\cite{Cuono2023EuInAs}. 

To second order in $\boldsymbol{p}$, the effective Hamiltonian term describing the altermagnet interaction is given by $\kappa p_x p_y \sigma_z$~\cite{Yuan2020altermagnetism}. Notice that even-in-momentum spin-dependent fields are possible in centrosymmetric crystals~\cite{Yuan2020altermagnetism,vsmejkal2022emerging}. In the following we will consider an altermagnet
\begin{equation}\label{eq:altermagnet_fields}
h_{\boldsymbol{n}}^z=\kappa p_F^2 n_x n_y.
\end{equation}
subjected to an in-plane Zeeman field $h_{\boldsymbol{n}}^x=h_x$. Other types of even-in-momentum fields have also been examined in works on WL of exciton-polaritons in microcavities with polarization splitting~\cite{Glazov2008Quantum,Glazov2010Spin}.

%
%
\begin{figure}[!t]
  \includegraphics[width=0.99\columnwidth]{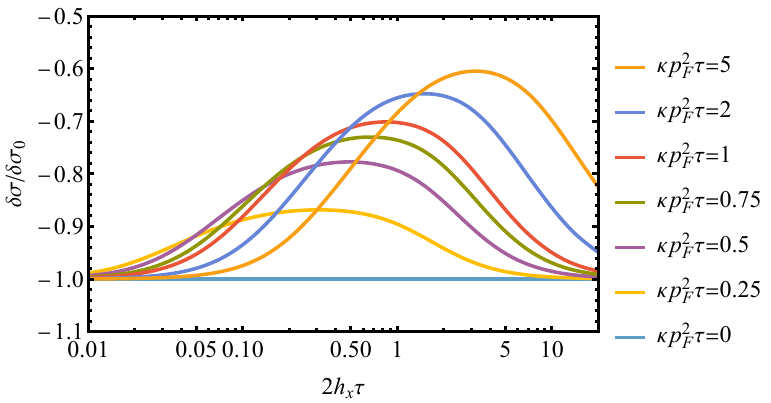}
\caption{Interference correction to the conductivity of an altermagnet as a function of the in-plane field $h_x$ for different values of the altermagnet spin splitting strength $\kappa$.}\label{Fig:conductance_vs_Zeeman_altermagnet}
\end{figure}
%
%

Inserting the values of the spin-dependent fields \eqref{eq:altermagnet_fields} into Eqs.~(\ref{eq:Pol}-\ref{eq:Wb}), we compute the WL conductivity of an altermagnet. In Fig.~\ref{Fig:conductance_vs_Zeeman_altermagnet} we show the WL conductivity as a function of the Zeeman field strength. In the absence of Zeeman field ($h_x=0$), the WL conductivity is not modified by the strength of the altermagnetism $\kappa$. Despite being a momentum-dependent interaction, the direction of the altermagnet term~\eqref{eq:altermagnet_fields} always lies along the $z$-direction. Unlike for Rashba or Dresselhaus type SOCs, there is a definite spin quantization axis which is independent of the crystal momentum. An incoming particle with up/down spin will not precess along any in-plane trajectory, so the SU(2) phase acquired by all paths contributing to the WL conductivity will be trivial. Therefore, the conductivity is not corrected by altermagnetism. When an inplane Zeeman field is applied, the spin quantization axis will depend on the momentum direction, so that spin is no longer an appropriate quantum number. This contrasts with the in-plane magnetoconductivity of a ferromagnetic material, where the total magnetic term $\boldsymbol{h}_{\boldsymbol{p}}$ does not depend on the momentum direction and thus does not modify the conductivity. For a very strong Zeeman field ($h_x \gg \kappa p_F^2$), a spin quantization axis along the $x$-direction is restored in the altermagnet, so the WL conductivity correction equals $\delta\sigma=-\delta \sigma_0$. The conductivity curves in Fig.~\ref{Fig:conductance_vs_Zeeman_altermagnet} display a bell-like shape. The value of the Zeeman field $h_x$ for which the maxima occur is slightly lower than $\kappa p_F^2$. Around this value, the interplay between the Zeeman and altermagnetic interactions is enhanced, resulting in a positive correction to the conductivity. This property might be useful to determine the strength of the altermagnetism $\kappa$ of a material through a WL measurement.

\section{Conclusion \label{Sec4}}
We have presented a theory of weak (anti)localization that accounts for any type of spin-dependent fields with arbitrary disorder strength. Equation~\eqref{eq:sigma} gives a very compact expression for the WL correction to the conductivity, which only requires the knowledge of the polarization operator $\hat{\Pi}_{\boldsymbol{nQ}}$ [Eq.~\eqref{eq:Pol}] to compute. This enables straightforward and efficient numerical computation of the WL correction. Moreover, in Sec.~\ref{Sec3C} we demonstrate that the same polarization operator also readily follows from the Eilenberger equation for superconducting systems, demonstrating the connection between WL and superconductivity. 

We expect that it is possible to derive a compact expression for the  WL correction similar to Eq.~$\eqref{eq:sigma}$ for any system where quasiclassical approximation holds. This includes systems with internal degrees of freedom other than spin (valleys, sublattice isospin, etc.), and systems with anisotropic disorder, or disorder which depends on the internal degrees of freedom. Similarly, we expect that the connection of WL and the Eilenberger equation also holds in general.

We apply our theory to calculate the WL conductivity in various systems. First, we test our theory on a system with Rashba and Dresselhaus SOC, which has been extensively studied in the context of WL in the past decades. Then, we apply our theory to more exotic systems -- the materials with Weyl-type SOC, and the newly discovered altermagnets.

We mainly focus on the in-plane magnetoconductivity, which shows distinct signatures created by the interplay of externally applied Zeeman field with the intrinsic spin-dependent fields of the material. Importantly, some of these signatures appear only beyond the conventionally used diffusive approximation. For instance, in Rashba systems the in-plane magnetoconductivity is non-monotonic as a function of field strength, while it oscillates a function of field direction. We show that it is possible to distinguish different kinds of linear-in-momentum SOC (Rashba, Dresselhaus and Weyl-type) in the non-diffusive regime, by measuring in-plane magnetoconductivity for different field directions. Finally, we discuss how to probe the strength of altermagnets by in-plane WL magnetoresistance measurements.

While in this work we have mainly focused on 2D systems, our results can also be applied to systems of any other dimensionality, e.g., quasi-1D nanowires with strong SOC. 

Our theory provides an efficient way to compute the WL conductivity, giving important insights for the classification of materials with SOC and other types of spin-dependent interactions. The simplicity of our expressions makes them very useful when fitting and interpreting experiments on systems with any type of spin texture, including traditional semiconducting structures, and novel van der Waals materials.

\begin{acknowledgments}
A.H. acknowledges funding from the University of the Basque Country (Project PIF20/05). F.S.B. and A.H. acknowledge financial support from Spanish MCIN/AEI/ 10.13039/501100011033 through project PID2020-114252GB-I00 (SPIRIT) and  TED2021-130292B-C42, and the Basque Government through grant IT-1591-22.  S.I. is supported by the Academy of Finland Research Fellowship (Project No. 355056).
\\

A.H. and S.I. contributed equally to this work.
\end{acknowledgments}

\appendix
\section{Derivation of main results from the diagrammatic theory \label{AppMainRes}}
In this Appendix we derive the results presented in Sec.~\ref{Sec2} from the standard diagrammatic theory. We first calculate the renormalized current vertex in Sec.~\ref{App1}, followed by the derivation of the polarization operator and the Cooperon in Sec.~\ref{App2}, and in Sec.~\ref{App3} we obtain the final result for the WL conductivity after evaluating the conductivity diagrams. 

To start, we introduce the disorder-averaged zero-temperature retarded (R) and advanced (A) Green's functions as
\begin{equation}
G_{\boldsymbol{p}}^{R,A}=\bigg(-\mathcal{H}_{0}\pm \frac{i}{2\tau}\bigg)^{-1},
\end{equation}
where $\mathcal{H}_0=\xi_{\boldsymbol{p}}+b_{\boldsymbol{n}}^i \sigma_i+h_{\boldsymbol{n}}^i \sigma_i$. 
Here, the self-energy $\pm i/(2\tau)$ is calculated in the self-consistent Born approximation. We assumed a random disorder potential $V$, characterized by the Gaussian white noise correlators $\langle V_{\boldsymbol{p}} \rangle_{\text{dis}}$=0 and $\langle V_{\boldsymbol{p}} V_{\boldsymbol{p'}}\rangle_{\text{dis}}=V_0^2 \delta_{\boldsymbol{p}\bar{\boldsymbol{p}}'}$, so that the inverse scattering time is $\tau^{-1}= 2\pi \nu V_0^2$. The brackets $\langle...\rangle_{\text{dis}}$ stand for disorder averaging. 

\subsection{Renormalized current vertex \label{App1}}
\begin{figure}[ht!]
\includegraphics[width=0.25 \textwidth]{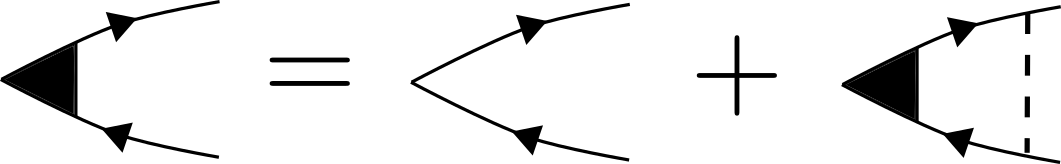}
\caption{Renormalization of the current vertex. Solid arrows represent Green's functions, while the dashed lines represent disorder. The upper (lower) branch of the diagrams correspond to retarded (advanced) Green's functions.}
\label{FigJ}
\end{figure}

The current operator is defined as $J_{i\boldsymbol{p}}=\partial_{p_i} \mathcal{H}_{\boldsymbol{p}}= v_F n_i+\partial_{p_i} b_{\boldsymbol{p}}^j\sigma_j+\partial_{p_i} h_{\boldsymbol{p}}^j\sigma_j$. In the leading order of the quasiclassical approximation, we need to keep only the first term and therefore $J_{i\boldsymbol{p}}\approx v_F n_i$. Next, we calculate the renormalized current vertex $\tilde{J}_{i\boldsymbol{p}}$, shown in the diagrammatic form in Fig.~\ref{FigJ}.

%
%
%
%

Explicitly writing the diagrams in Fig.~\ref{FigJ}, we obtain
\begin{equation}
\tilde{J}_{i\boldsymbol{p}}=J_{i\boldsymbol{p}}+\frac{1}{2\pi \tau \nu}\int \frac{d^d \boldsymbol{p'}}{(2\pi)^d} G_{\boldsymbol{p'}}^A \tilde{J}_{i\boldsymbol{p'}} G^R_{\boldsymbol{p'}}.
\end{equation}
We look for the solution of this equation in the form $\tilde{J}_{i\boldsymbol{p}}=K J_{i\boldsymbol{p}}=K v_F n_i$, where $K$ is a scalar. The momentum integration is performed by substitution $\int \frac{d^d \boldsymbol{p}}{(2\pi)^d} \ldots \to \nu \langle \int d\xi_\mathbf{p} \ldots \rangle$ and using the residue theorem. Performing the energy integral yields $\nu \int d\xi_{\boldsymbol{p}} G_{\boldsymbol{p}}^A G_{\boldsymbol{p}}^R=2\pi \tau \nu$. Then, the equation for the renormalized vertex reduces to
\begin{equation}
K v_F n_i=v_F n_i +\langle K v_F n_i \rangle =v_F n_i.
\end{equation}
Therefore, $K=1$, and the renormalized current vertex is $\tilde{J}_{i\boldsymbol{p}}=J_{i\boldsymbol{p}}=v_F n_i$.

\subsection{Polarization operator and the Cooperon \label{App2}}
\begin{figure}[ht!]
\includegraphics[width=0.45 \textwidth]{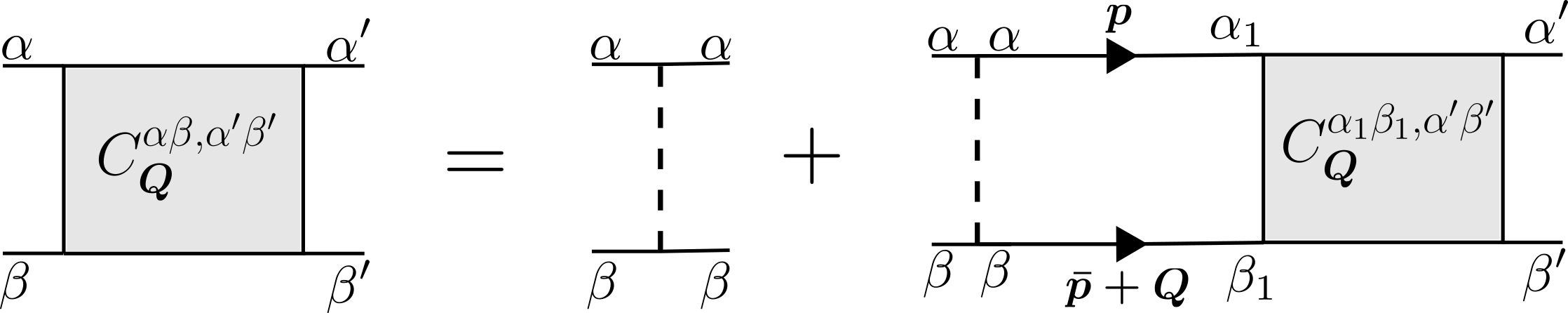}
\caption{Bethe-Salpeter equation for the Cooperons. Greek indices in the subscript denote the spin. Summation over repeated indices is assumed.}
\label{FigC}
\end{figure}

The central object in the theory of WL is the Cooperon, which is determined from the Bethe-Salpeter equation, shown in the diagrammatic form in Fig.~\ref{FigC}.
%
%
%
%

Explicitly writing the diagrams from Fig.~\ref{FigC} yields
\begin{equation}
C_{\boldsymbol{Q}}^{\alpha\beta,\alpha'\beta'}=\frac{1}{2\pi \tau \nu} \delta_{\alpha\alpha'}\delta_{\beta\beta'} +\Pi_{\boldsymbol{Q}}^{\alpha\beta,\alpha_1\beta_1} C_{\boldsymbol{Q}}^{\alpha_1\beta_1,\alpha'\beta'}.
\end{equation}
The Greek indices denote the spin degree of freedom, and can take value $\uparrow$ or $\downarrow$. The summation over repeated indices is assumed. 
We introduced the angle-averaged polarization operator 
\begin{equation}
\Pi_{\boldsymbol{Q}}^{\alpha\beta,\alpha_1\beta_1}= \frac{1}{2\pi \tau\nu} \int \frac{d^d\boldsymbol{p}}{(2\pi)^d} [G^R_{\boldsymbol{p}}]_{\alpha\alpha_1}[G^A_{\bar{\boldsymbol{p}}+\boldsymbol{Q}}]_{\beta\beta_1}, 
\end{equation}
where  we introduced the notation $\bar{\boldsymbol{p}}=-\boldsymbol{p}$.

Next, we transform the polarization operator and the Cooperon to the singlet-triplet basis using the transformation 
\begin{equation}
M^{ss'}=\frac{1}{2} [\sigma_y\sigma_s]_{\alpha\beta}M^{\alpha\beta,\alpha'\beta'}
[\sigma_{s'}\sigma_y]_{\beta'\alpha'}. 
\end{equation}
The inverse transformation is
\begin{equation}
M^{\alpha\beta,\alpha'\beta'}=\frac{1}{2} [\sigma_s \sigma_y]_{\beta\alpha} M^{ss'} [\sigma_y \sigma_{s'}]_{\alpha'\beta'}.
\end{equation} 
Here, the indices $s,s'=0,x,y,z$ label the singlet $0$ and the triplets $x,y,z$. 
After the transformation, we obtain
\begin{equation}
C^{ss'}_{\boldsymbol{Q}}=\frac{1}{2\pi \nu \tau} \delta_{ss'}+\Pi^{ss_1}_{\boldsymbol{Q}}C^{s_1s'}_{\boldsymbol{Q}},
\end{equation}
or, in matrix form
\begin{equation}
\hat{C}_{\boldsymbol{Q}}=\frac{1}{2\pi \tau \nu}+\hat{\Pi}_{\boldsymbol{Q}}\hat{C}_{\boldsymbol{Q}}.
\end{equation}
Inverting this equation to express $\hat{C}$ yields Eq.~\eqref{Eq:Coop} in the main text. The polarization operator in the singlet-triplet basis is $\Pi^{ss'}_{\boldsymbol{Q}}=\langle  \Pi^{ss'}_{\boldsymbol{n}\boldsymbol{Q}}\rangle$, where 
\begin{equation}
\Pi^{ss'}_{\boldsymbol{n}\boldsymbol{Q}}=\frac{1}{4\pi \tau} \int d\xi_{\boldsymbol{p}} \text{Tr}[\sigma_y \sigma_s G^{A}_{\bar{\boldsymbol{p}}+\boldsymbol{Q}}\sigma_{s'}\sigma_y (G^R_{\boldsymbol{p}})^\top]. 
\label{Eq:PolInt}
\end{equation}
The integration in Eq.~\eqref{Eq:PolInt} can be performed for each pair of indices $s$ and $s'$ by the residue theorem, which we do using \emph{Mathematica}.  This finally yields Eq.~\eqref{eq:Pol} of the main text.

\begin{table*}[!t]
\centering
\begin{tabular}{ | c | c | c | } 
  \hline
  System 1 & System 2 & Symmetry operation \\ 
  \hline
  Rashba ($\alpha$) + Dresselhaus ($\beta$) & Rashba ($\beta$) + Dresselhaus ($\alpha$) & $\widehat{R}_{\hat{x}}(\pi)\widehat{R}_{\hat{z}}(\pi/2)$ \\ 
  \hline
  Rashba + Zeeman field $\boldsymbol{h}=(h_x,h_y,0)$ & Dresselhaus + Zeeman field $\boldsymbol{h}=(-h_y,-h_x,0)$ & $\widehat{R}_{\hat{x}}(\pi)\widehat{R}_{\hat{z}}(\pi/2)$ \\ 
  \hline
  Rashba + Zeeman field $\boldsymbol{h}=(h_x,h_y,0)$ & Weyl-type SOC + Zeeman field $\boldsymbol{h}=(-h_y,h_x,0)$ & $\widehat{R}_{\hat{z}}(\pi/2)$ \\ 
  \hline
  Dresselhaus + Zeeman field $\boldsymbol{h}=(h_x,h_y,0)$ & Weyl-type SOC + Zeeman field $\boldsymbol{h}=(h_x,-h_y,0)$ & $\widehat{R}_{\hat{x}}(\pi)$ \\ 
  \hline
\end{tabular}
\caption{Systems with equivalent WL correction and the spin symmetry operation relating them. $\widehat{R}_{\hat{\boldsymbol{n}}}(\varphi)$ is a $\varphi$ proper rotation around the $\hat{\boldsymbol{n}}$ axis. Lines 2-4 show how the different types of linear-in-momentum SOCs with a Zeeman field are related to each other.}
\label{table:spin_symmetries}
\end{table*}

\subsection{WL corrections to the conductivity \label{App3}}
\begin{figure}[ht!]
    \centering
    \includegraphics[width=0.45\textwidth]{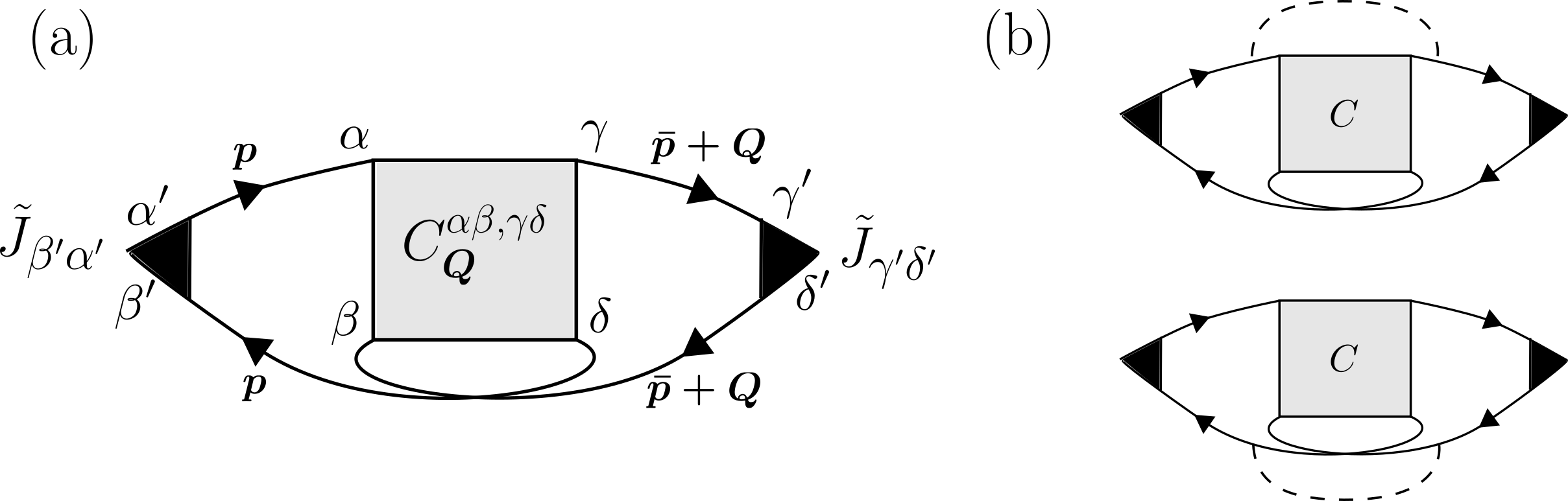}
    \caption{Diagrams for the WL correction to the conductivity. (a) Bare Hikami box. (b) Dressed Hikami boxes. }
    \label{Fig1}
\end{figure}

Diagrammatic representation of the WL correction to the conductivity is given in Fig.~\ref{Fig1}. We distinguish two contributions to the WL conductivity
\begin{equation}
\delta\sigma=\delta\sigma^{(a)}+\delta \sigma^{(b)},
\end{equation}
where $\delta\sigma^{(a)}$ in the backscaterring contribution coming from the so-called bare Hikami boxes [Fig.~\ref{Fig1}(a)], and $\delta\sigma^{(b)}$ is the non-backscattering contribution \cite{dmitriev1997nonbackscattering} coming from the dressed Hikami boxes [Fig.~\ref{Fig1}(b)].

%
%
%
%

Explicitly writing the diagrams in Fig.~\ref{Fig1}(a) while taking that the current flows along the $x$-direction yields
\begin{multline}
\delta \sigma^{(a)}=\frac{e^2}{2\pi}\int \frac{d^d\boldsymbol{p}}{(2\pi)^d}\int \frac{d^d\boldsymbol{Q}}{(2\pi)^d} [G^R_{\boldsymbol{p}}]_{\alpha'\alpha}C^{(a)\alpha\beta,\gamma\delta}_{\boldsymbol{Q}} \\ 
\times [G_{\boldsymbol{\bar{p}+Q}}^R]_{\gamma\gamma'} 
[\tilde{J}_{x\boldsymbol{\bar{p}}}]_{\gamma'\delta'} [G_{\boldsymbol{\bar{p}+Q}}^A]_{\delta'\beta}[G_{\boldsymbol{p}}^A]_{\delta\beta'}[\tilde{J}_{x\boldsymbol{p}}]_{\beta'\alpha'},
\label{eqa4}
\end{multline}
 Note that conductivity diagrams with single and two impurity lines diverge upon $Q$ integration \cite{gasparyan1985field, cassam1994two, dyakonov1994magnetoconductance, dmitriev1997nonbackscattering, zduniak1997universal, golub2005weak,glazov2006nondiffusive,glazov2009spin}. For this reason, we introduce the reduced Cooperon $C^{(a)}$, where the contributions from single and two impurity lines are removed. Namely
\begin{equation}
C_{\boldsymbol{Q}}^{(a)\alpha\beta,\gamma\delta}=C_{\boldsymbol{Q}}^{\alpha\beta,\gamma\delta}-\frac{1}{2\pi \nu \tau }[\delta_{\alpha\beta}\delta_{\gamma\delta}+\Pi_{\boldsymbol{Q}}^{\alpha\beta,\gamma \delta}].
\label{Eq:Coopa}
\end{equation}
Transforming Eq.~\eqref{Eq:Coopa} to the singlet-triplet basis, we obtain
\begin{equation}
\hat{C}^{(a)}_{\boldsymbol{Q}}=\hat{C}_{\boldsymbol{Q}}-\frac{1}{2\pi \nu\tau}(1+\hat{\Pi}_{\boldsymbol{Q}})=\hat{C}_{\boldsymbol{Q}}\hat{\Pi}_{\boldsymbol{Q}}^2.
\end{equation}
Next, transforming Eq.~\eqref{eqa4} to the singlet-triplet basis yields
\begin{equation}
\delta\sigma^{(a)}=\frac{e^2}{2\pi}\int \frac{d^d\boldsymbol{Q}}{(2\pi)^d}\text{Tr}[\hat{C}_{\boldsymbol{Q}}^{(a)} \hat{W}^{(a)}_{\boldsymbol{Q}}],
\label{eqa6}
\end{equation}
where we introduced the weight matrix  $\hat{W}^{(a)}_{\boldsymbol{Q}}$ as
\begin{multline}
W_{\boldsymbol{Q}}^{(a)ss'}=\frac{1}{2} \int \frac{d^d\boldsymbol{p}}{(2\pi)^d}\\
\text{Tr}\bigg[\sigma_y\sigma_{s'} G^A_{\boldsymbol{p}}\tilde{J}_{x\boldsymbol{p}}G^R_{\boldsymbol{p}} \bigg( G^R_{\boldsymbol{\bar{p}}+\boldsymbol{Q}}\tilde{J}_{x\boldsymbol{\bar{p}}}G^A_{\boldsymbol{\bar{p}}+\boldsymbol{Q}} \sigma_y\sigma_{s}\bigg)^\top \bigg].
\label{eqa7}
\end{multline}
After performing $\xi_{\boldsymbol{p}}$-integration by residue theorem using \emph{Mathematica}, we obtain Eq.~\eqref{eq:Wa} from the main text.

The non-backscattering WL conductivity $\delta \sigma^{(b)}$ is similarly obtained as
\begin{equation}
\delta\sigma^{(b)}=\frac{e^2}{2\pi}\int \frac{d^d\boldsymbol{Q}}{(2\pi)^d}\text{Tr}[\hat{C}_{\boldsymbol{Q}}^{(b)} \hat{W}^{(b)}_{\boldsymbol{Q}}].
\label{eq:sigmab}
\end{equation}
Here, in the reduced Cooperon $C^{(b)}$ we need to subtract the single-impurity contribution (it ultimately gives a two-impurity conductivity diagram, since the weight factor already contains one impurity line):
\begin{equation}
\hat{C}_{\boldsymbol{Q}}^{(b)}=\hat{C}_{\boldsymbol{Q}}-\frac{1}{2\pi\nu \tau}=\hat{C}_{\boldsymbol{Q}}\hat{\Pi}_{\boldsymbol{Q}}.
\end{equation}
The weight matrix $\hat{W}_{\boldsymbol{Q}}^{(b)}$ is determined from 
\begin{multline}
W^{(b)ss'}_{\boldsymbol{Q}}=\frac{1}{2} \frac{1}{2\pi \nu \tau}\int \frac{d^d\boldsymbol{p}}{(2\pi)^d} \int \frac{d^d\boldsymbol{p'}}{(2\pi)^d} \text{Tr}\bigg[ \\
\sigma_y\sigma_{s'} G^A_{\boldsymbol{p}}\tilde{J}_{x\boldsymbol{p}}G^R_{\boldsymbol{p}}G^R_{\boldsymbol{p'}} \bigg(G^R_{\boldsymbol{\bar{p}}+\boldsymbol{Q}} G^R_{\boldsymbol{\bar{p}}'+\boldsymbol{Q}}\tilde{J}_{x\boldsymbol{\bar{p}'}}G^A_{\boldsymbol{\bar{p}'+Q}} \sigma_y\sigma_{s}\bigg)^\top \\
+\sigma_y\sigma_{s'} G^A_{\boldsymbol{p'}}G^A_{\boldsymbol{p}}\tilde{J}_{x\boldsymbol{p}}G^R_{\boldsymbol{p}} \bigg( G^R_{\boldsymbol{\bar{p}'}}\tilde{J}_{x\boldsymbol{\bar{p}'}}G^A_{\boldsymbol{\bar{p}'}+\boldsymbol{Q}}G^A_{\boldsymbol{\bar{p}+Q}} \sigma_y\sigma_{s}\bigg)^\top\bigg].
\label{eqa8}
\end{multline}
Here, the second and third line come from the two different types of dressed Hikami boxes, represented in the upper and lower panels of Fig.~\ref{Fig1}(b), respectively. Evaluating the integrals over $\xi_{\boldsymbol{p}}$ and $\xi_{\boldsymbol{p'}}$ by residue theorem in \emph{Mathematica} yields Eq.~\eqref{eq:Wb} of the main text.

Finally, the total WL correction is given by summing Eq.~\eqref{eqa6} and Eq.~\eqref{eq:sigmab}, giving  Eq.~\eqref{eq:sigma} from the main text.

\section{Polarization operator in the diffusive limit for systems with Rashba and Dresselhaus SOC \label{AppDif}}
In this Appendix, we explicitly write the polarization operator in the diffusive limit for the widely studied 2D system with Rashba and Dresselhaus SOC. Using Eq.~\eqref{Eq:PolDif} together with Eqs.~\eqref{Eq:Rashba} and \eqref{eq:Zeeman} gives
\begin{widetext}
\begin{equation}
\label{eq:diffusive_limimt}
1-\hat{\Pi}_{\boldsymbol{Q}}= l/l_\phi+D Q^2 \tau+ 
\begin{pmatrix}
\Gamma_{DP}\tau +x_{hx}^2 & x_{hx}x_{hy}+x_{\alpha}x_{\beta1} & -i (x_\alpha l Q_x+x_{\beta1} l Q_y) & -i x_{hx} \\
x_{hx}x_{hy}+x_{\alpha}x_{\beta1} & \Gamma_{DP}\tau+x_{hy}^2 & -i (x_{\beta1} l Q_x+x_{\alpha} l Q_y) & - i x_{hy} \\
i (x_\alpha l Q_x+x_{\beta1} l Q_y) & i (x_{\beta1} l Q_x+x_{\alpha} l Q_y) & 2 \Gamma_{DP}\tau & 0\\
-i x_{hx} & -i x_{hy} & 0 & x_{hx}^2+x_{hy}^2
\end{pmatrix}.
\end{equation}
\end{widetext}
Here, we introduced the dimensionless parameters quantifying the magnitude of the different spin-fields: $x_\alpha=2\alpha p_F \tau$, $x_{\beta 1}=2\beta p_F\tau$, $x_{\beta3}=2 \beta_3 p_F^3 \tau$, $x_{hx}=2 h_x \tau$ and $x_{hy}=2 h_y \tau$. $\Gamma_{DP}=(x_\alpha^2+x_{\beta1}^2+x_{\beta3}^2)/2\tau$ is the D'yakonov-Perel spin relaxation rate, and $\boldsymbol{Q}=(Q_x,Q_y)$. The polarization operator~\eqref{eq:diffusive_limimt} recovers known results ~\cite{iordanskii1994weak, pikus1995conduction, knap1996weak, punnoose2006magnetoconductivity, marinescu2019closed, weigele2020symmetry, mal1997magnetoresistance, marinescu2006electron, marinescu2017cubic}.

\section{Spin rotation of the spin-dependent fields\label{AppRot}}
In this appendix, we show that systems where the spin-dependent fields are related through spin symmetry operations give rise to the same WL correction to the conductivity. A comprehensive symmetry analysis for systems with Zeeman and linear-in-momentum SOC was developed in Ref.~\cite{Aleiner2001SOC} in the diffusive regime. Here we account for spin-dependent fields with arbitrary momentum dependence. If the Zeeman-like fields $\boldsymbol{h}_{\boldsymbol{n}}=(h_{\boldsymbol{n}}^x,h_{\boldsymbol{n}}^y,h_{\boldsymbol{n}}^z)$ and the SOC-related fields $\boldsymbol{b}_{\boldsymbol{n}}=(b_{\boldsymbol{n}}^x,b_{\boldsymbol{n}}^y,b_{\boldsymbol{n}}^z)$ in the Hamiltonian~\eqref{Hamiltonian} are transformed through pure spin operations, i.e. distance-preserving transformations in spin-space without transforming the spatial degrees of freedom, the resulting integrand in Eq.~\eqref{eq:sigma} is equivalent to the original one, so both systems have equivalent WL correction to the conductivity.

Within our formalism, the $4 \times 4$ matrices in singlet/triplet space transform as $\hat{M} \rightarrow \hat{U}\hat{M}\hat{U}^\top$, where the symmetry operation matrix $\hat{U}$ is given by
\begin{equation}
    \hat{U}=\begin{pmatrix}
        \widehat{R} & \boldsymbol{0}\\
        \boldsymbol{0} & 1
    \end{pmatrix}\; ,
\end{equation}
where $\widehat{R}$ is a $3 \times 3$ orthogonal matrix describing proper and improper rotations. Inserting the identity $\hat{U}\hat{U}^\top$ in the trace in Eq.~\eqref{eq:sigma}, we obtain
\begin{equation}
\text{Tr}\left[\hat{C}\hat{W}\right]=\text{Tr}\left[\hat{U}\hat{C}\hat{U}^\top\hat{U}\hat{W}\hat{U}^\top\right]\; .
\end{equation}
The Cooperon~\eqref{Eq:Coop} transforms as
\begin{equation}
    \hat{U}\hat{C}\hat{U}^\top=\frac{1}{2\pi \nu \tau}\hat{U}[1-\hat{\Pi}_{\boldsymbol{Q}}]^{-1}\hat{U}^\top=\frac{1}{2\pi \nu \tau}[1-\hat{U}\hat{\Pi}_{\boldsymbol{Q}}\hat{U}^\top]^{-1}\; ,
\end{equation}
and similarly the polarization operator~\eqref{eq:Pol} transforms as
\begin{equation}
\hat{U}\hat{\Pi}_{\boldsymbol{n}\boldsymbol{Q}}\hat{U}^\top= (1+l/l_\phi+i l \boldsymbol{n}\cdot \boldsymbol{Q}+\hat{U}\hat{A}_{\boldsymbol{n}}\hat{U}^\top)^{-1}.
\end{equation}
Following a similar procedure for the weighting factor, one also obtains that transforming $\hat{W}$ is equivalent to transforming the $\hat{A}_{\boldsymbol{n}}$ matrix inside of it.

Using the explicit form of matrix $\hat{A}_{\boldsymbol{n}}$~\eqref{eq:An}, it can be shown that the magnetic terms $\boldsymbol{h}_{\boldsymbol{n}}$ and the SOC terms $\boldsymbol{b}_{\boldsymbol{n}}$ transform as vectors and pseudovectors under the symmetry operation $\widehat{R}$, respectively. This means that under spin rotations, the spin-dependent fields transform as $\boldsymbol{h}_{\boldsymbol{n}} \rightarrow \widehat{R}\boldsymbol{h}_{\boldsymbol{n}}$ and $\boldsymbol{b}_{\boldsymbol{n}} \rightarrow \widehat{R}\boldsymbol{b}_{\boldsymbol{n}}$ under proper rotations and $\boldsymbol{h}_{\boldsymbol{n}} \rightarrow \widehat{R}\boldsymbol{h}_{\boldsymbol{n}}$ and $\boldsymbol{b}_{\boldsymbol{n}} \rightarrow -\widehat{R}\boldsymbol{b}_{\boldsymbol{n}}$ under improper rotations. In conclusion, the integrand in Eq.~\eqref{eq:sigma} for two systems whose spin-dependent fields are related through spin symmetry operations is the same, so both systems have equivalent WL correction to the conductivity.

In the following, we provide a simple example of two systems that give rise to the same WL conductivity. The spin-fields for a system with only Rashba SOC are $b_{\boldsymbol{n}}^x=\alpha p_F n_y$ and $b_{\boldsymbol{n}}^y=-\alpha p_F n_x$. Rotating $\boldsymbol{b}_{\boldsymbol{n}}$ by $\pi/2$ around the $\boldsymbol{z}$ axis followed by a $\pi$ rotation around the $\boldsymbol{x}$ axis results in a spin-field $\tilde{b}_{\boldsymbol{n}}^x=\alpha p_F n_x$ and $\tilde{b}_{\boldsymbol{n}}^y=-\alpha p_F n_y$, which is equivalent to a linear Dresselhaus SOC with strength $\alpha$. More generally, the WL conductivity of a system subjected to Rashba and linear Dressehaus SOC is the same as another system with interchanged SOC strengths $\alpha \leftrightarrow \beta$. This SOC induced localization symmetry has already been reported in quasi-1D quantum networks~\cite{Ramaglia2006symmetry,Hijano2023spintopology}. In Table~\ref{table:spin_symmetries} we present other examples of systems with equivalent WL correction and the spin symmetry operations relating them.

%

\end{document}